# Insights of Dielectric relaxations in Nd and Mn co-substituted BiFeO$_3$

Chandrakanta Panda[1], Pawan Kumar[1,2], A. K. Sinha[3], Archna Sagdeo[3] and Manoranjan Kar[1*]

[1]Department of Physics, Indian Institute of Technology Patna, Bihta-801103, India.

[2]Department of Physics, National Institute of Technology, Jamshedpur -831014, India

[3]Indus Synchrotron Utilization Division, RRCAT Indore, Indore-452013, India

* Corresponding author Email: mano@iitp.ac.in, Ph: +916122552013, Fax: +916122277383

**Abstract:**

The composition-driven structural transition from the rhombohedral (*R*3*c*) to orthorhombic (*Pbnm*) symmetry is observed from crystallographic phase analysis of Synchrotron X-ray diffraction patterns. Highest dielectric constant is observed for Bi$_{0.95}$Nd$_{0.05}$Fe$_{0.95}$Mn$_{0.05}$O$_3$ (BNFM-05) sample where both crystallographic phases are present with equal percentage. An integral analysis and discussion on temperature-dependent impedance spectroscopy have been carried out to understand the dielectric relaxations in BNFM-05 ceramics. The ac conductivity data obeys Jonscher's power law and is consistent with the small polaron hopping conduction mechanism. The bulk electrical conductivity suggests an Arrhenius type of thermally activated process with three different conduction mechanisms. It indicates the non-Debye type of dielectric relaxation. The dielectric constant dispersion in the frequency range of 20 Hz –1 MHz has been analyzed by employing the modified Debye's function with multiple charge carriers contributing to the relaxation. The dielectric anomaly at the magnetic transition temperature i.e. antiferromagnetic Neel temperature ($T_N$) has been observed which was predicted on the basis of Landau theory in multiferroic materials with magnetoelectric coupling.

## 1. Introduction

The coexistence of (anti)ferroelectric and (anti)ferromagnetic orders in a single material have attracted the attention of several research groups due to their potential applications in memory devices, magnetically modulated transducers, spintronic devices, ultrafast optoelectronic devices and sensors etc [1-4]. BiFeO$_3$ (BFO) is the most promising multiferroics because of ferroelectric as well as G-type antiferromagnetic orderings above room temperature ($T_C$ = 830 °C and $T_N$ = 370 °C) which promises the possibility of developing potential devices based on magnetoelectric coupling operating at room temperature [5]. Its crystal structure is described by rhombohederally distorted perovskite structure with the *R*3*c* space group. The ferroelectricity in this compound arises due to the off-centre structural distortions of cations induced by the 6S$^2$ lone pair electrons of Bi$^{3+}$ whereas the G-type antiferromagnetic ordering arises due to indirect exchange interactions between Fe$^{3+}$ ions through O$^{2-}$ ions [6]. The technological applications of BFO are limited due to high leakage current which originates from oxygen off-stoichiometry, high defect density and impurity phases (Bi$_2$Fe$_4$O$_9$ and Bi$_{25}$FeO$_{40}$ etc.). Oxygen vacancies in BFO are due to the inter-conversion from Fe$^{3+}$ to Fe$^{2+}$.

The synthesis of single phase BFO is difficult because the constituent oxides like Bi$_2$O$_3$ and Fe$_2$O$_3$ have adverse thermodynamic characters as Bi$_2$O$_3$ is volatile and Fe$_2$O$_3$ is less reactive at low temperature. Therefore, the synthesis is limited within narrow temperature range. The chemical route of synthesis has been adopted for uniform particle size and better reproducibility. As the ferroelectricity in BFO is due to the presence of $6S^2$ lone pair in Bi cations, several attempts have been made to create chemical pressure at the lattice sites by different substituents at Bi as well as Fe sites to create large displacement change at Bi sites and hence enhance the ferroelectric properties. Hence, the stress has been created at the both lattice sites by the substitution of Nd and Mn in Bi and Fe sites respectively. Recently it has been reported that the co-substitution of Nd and Mn shows enhanced magnetization for 5 % of co-substitution where the maximum phase competition between two crystal symmetries such as rhombohedral and orthorhombic has been observed [7]. The high ferroelectric and ferromagnetic polarization at room temperature has been reported in $Nd^{3+}$ and $Mn^{3+}$ substituted BFO [8]. However, the details of dielectric relaxations for the composition with equal percentage of both crystallographic phases (Rhombohedral and orthorhombic) are present where highest dielectric constant has been observed. Hence, it will be of considerable interest to investigate the details about the dielectric properties for this particular composition which will explore the physics behind the dielectric properties which will open the window to tune the dielectric constant for this promising multiferroic material. In the present work, a systematic study of temperature variation complex impedance spectroscopy has been carried out to understand dielectric relaxations of Nd and Mn co-substituted BFO ceramics as well as structural transition have been addressed by Synchrotron diffraction technique.

## 2. Materials and Methods

$Bi_{1-x}Nd_xFe_{1-x}Mn_xO_3$ with x = 0.025, 0.050, 0.075 and 0.100 (named as BNFM-025, BNFM-05, BNFM-075 and BNFM-10 respectively) were synthesized by the tartaric acid modified sol-gel technique [7]. The resulting materials were annealed at 700 $^oC$ for 3 hours. The compositional-driven crystallographic phase transition was studied by employing the X-ray diffraction technique and details were reported in our earlier publication [7]. However, the high power Synchrotron XRD measurements were performed at angle dispersive X-ray diffraction (ADXRD) beamline (BL-12) [9] on Indus-2 synchrotron source [10] to confirm the compositional-driven crystallographic phase transition. The X-ray wavelength used for the present study was accurately calibrated by doing X-ray diffraction on $LaB_6$ (NIST standard). Data processing was done using Fit2D software [11]. The parallel plane surfaces of pellets were coated with the silver paste for dielectric measurements. The impedance measurements were performed using the impedance analyzer (N4L PSM 1735) in the frequency range of 300 Hz – 8 MHz and temperature range of 30 - 600 $^oC$. Synchrotron XRD patterns were analyzed by employing the Rietveld refinement technique with the help of Fullprof software package to extract the crystallographic parameters [12, 13]. Final refined structural parameters for all samples are mentioned in the Table 1. All the occupancy parameters were taken as fixed at their corresponding composition during refinement process. The peaks corresponding to impurity phase identified as $Bi_{25}FeO_{40}$ were not considered during the Rietveld analysis.

## 3. Results

Synchrotron X-Ray diffraction (SXRD) patterns for all $Bi_{1-x}Nd_xFe_{1-x}Mn_xO_3$ ceramics annealed at 700 $^oC$ have been shown in the Fig. 1. Small impurity phases as $Bi_{25}FeO_{40}$ and $Bi_2Fe_4O_9$ have been observed from SXRD pattern. These impurities phases are substantially suppressed with the increase in co-substitution of Nd & Mn. However, the peak intensity is very small for the BLFM-05 sample where both crystallographic phases are present in equal percentage as analyzed by the Rietveld analysis and discussed below. Another observation is that the only $Bi_{25}FeO_{40}$ impurity is present for the samples with x > 0.05. The (104) and (110) reflections were clearly separated for lower co-substitution concentration. With the increase in the substituents concentration, all the doublets appears merging to give a single peak, which is clearly visible for more than 5 % of co-substitution. These results correspond to the decrease in the crystallite size and increase in lattice distortion for co-substituted ceramics. The lattice constants (Table 1) decrease due to smaller radius of substituents than that of host cations, which is also evident from the shift of the XRD peak towards higher $2\theta$ value (as shown in Fig. 2). This shift indicates the incorporation of substituted cations in the BFO lattice. The crystallite size has been calculated using Scherrer's formula [14]. The decrease in crystallite size implies the development of lattice strain due to ionic size mismatch between $Nd^{3+}$ & $Bi^{3+}$ and $Fe^{3+}$ & $Mn^{3+}$. It leads to local structural disorder and reduces the rate of nucleation as an effect decrease the crystallite size. The Rietveld analysis of all the SXRD patterns indicates the composition driven structural transition from the rhombohedral ($R3c$) to orthorhombic ($Pbnm$) symmetry. The typical Rietveld refined XRD patterns for the BNFM-05 and BNFM-075 have been shown in the Fig. 3. The structural parameters and their variation with composition are similar to that of report in the literature for lab source XRD [7]. The weight percentage of the orthorhombic crystallographic phase increases with the increase in the substitution and it becomes the dominant crystal phase for the co-substitution with x > 0.05.

The real part of impedance (Z') versus frequency plot of BNFM-05 ceramics at different temperatures (40, 100, 150, 200 and 250 $^oC$) have been shown in the Fig. 4. Presence of dielectric relaxation behaviour is evident from the dispersion behaviour of Z' in the low frequency regions. The decrease in Z' with the increase in frequency is due to the increase in hopping of charge carrier which leads to the increase in ac conductivity. The increase in Z' with the increase in Nd and Mn substitution indicates that the bulk resistance was enhanced upon co-substitution. It may be due to increase in the effective potential barrier heights with the structural distortions as discussed above using synchrotron X-Ray diffraction analysis. The lattice strain is developed due to ionic size mismatch of host and substituting cations leading to the structural distortion. All the samples exhibit negative temperature coefficient of resistance (NTCR) character i.e. decrease in Z' with the increase in temperature. Fig. 5 shows frequency dependence of the imaginary part of impedance ($Z''$) of BNFM-05 ceramics. It has different characteristics such as (i) appearance of a peak with a maximum value ($Z''_{max}$) at a certain frequency, (ii) asymmetrical broadening, (iii) shifting of $Z''_{max}$ value to higher frequency with the increase in temperature, and (iv) decrement of magnitude of $Z''_{max}$ with the temperature. Appearance of peak suggests the presence of dielectric relaxation. Asymmetric broadening of peak indicates the distribution of relaxation times which refers to the non-Debye behaviour.

Generally, polycrystalline dielectric materials have grains separated by interfacial boundary layers (grain boundaries). The corresponding features can be observed with two semi-circular arcs in the Niquist plots. The low frequency semi-circular arc corresponds to the grain boundary and high frequency semicircular arc to the grain. If there exits multi-relaxations in the material, the semicircular arcs are depressed, i.e., dielectric relaxation deviates from ideal Debye relaxation behaviour. Each semi-circular arc can be represented as an equivalent electrical circuit consisting of a parallel combination of a resistor R and a capacitor *C* (or constant phase element CPE) as shown in the Fig. 6. The reliable values of resistance of grain and grain boundary have been obtained by analyzing with the help of equivalent circuit model (i.e. brick layer model) shown in the Fig. 6 [15]. In the Fig. 6, $R_s$ is the resistance of measuring leads, $C_g$ is the capacitance due to domain and dipole reorientation in grain, $R_g$ is the resistance of grain, $C_{gb}$ is the capacitance in the grain boundary layer, $R_{gb}$ denotes the resistance across the grain boundary layer, and CPE is a constant phase element which indicates the departure from ideal Debye-type model [16]. The CPE admittance is defined as follows,

$Y_{CPE} = A\omega^n + jB\omega^n$ - - - - - - - - - - - - (1)

$A = A_0 \cos\left(\frac{n\pi}{2}\right)$ and $B = B_0 \sin\left(\frac{n\pi}{2}\right)$ - - - - - - - - - - - - (2)

The parameters '$A_0$' eq$^n$ (2) defines the magnitude of the dispersion and '$n$' in eq$^n$ (1) varies as $0 \leq n \leq 1$. The parameter '$n$' is equal to 1 and 0 for ideal capacitor and ideal resistor respectively [16]. The experimental data were then fitted using '*ZSIMPWIN*' software by using the equivalent circuit as discussed above. $R_s$, $R_g$, $C_g$, $R_{gb}$ and $C_{gb}$ thus obtained are summarized in the Table 2. The complex impedance data obtained for BNFM-05 ceramics at various temperatures ranging from 40 to 250 °C have been shown as Nyquist plots in the Fig. 7. The straight lines with large slope appear at low temperature which indicates that the sample is highly insulating. However, the impedance decreases with the increase in temperature. The two semicircular arcs are poorly resolved which are due to the responses of grains and that of grain boundaries respectively. The small arc at high frequencies tends is not clearly observable due to large arc at low frequencies which may be due to the large difference in the magnitudes of $R_g$ and $R_{gb}$. As the temperature increases, the circular arc depresses as shown in the Fig. 7, and the intercept of arc with the *Z'* axis shifts towards the origin in the Nyquist plots which indicates the decrease in bulk resistance of the compound. It indicates that the present samples showed NTCR character which is consistent with the above discussion.

Since the relaxation peak at lower frequency (<10$^4$ Hz) can be better analyzed by conductivity loss, hence the investigation into the electric modulus (M*) was conducted which is defined as follows;

M* = 1/ε* = M' + jM" = jωC$_0$Z* = jωC$_0$ (Z' − jZ") - - - - - - - - - - - - (3)

As electric modulus is the inverse of complex permittivity, more conductivity loss contributes to the dielectric permittivity but less conductivity loss affects the modulus. M' & M" versus frequency plots in the temperature range of 40 - 250 °C are shown in Figs. 8 & 9. One can observe that the value of M' increases with the increase in frequency. As there is only one plateau and peak in the M' and M" curve respectively, it indicates that there is only one thermally activated process in the present measurement range. M' shows the dispersion which increases with the increase in frequency and shifted towards the higher frequency side with the increase in temperature. Frequency variation of M" curves are characterized by (i) appearance of peak at unique frequency, at a given temperature, (ii) significant broadening in the peak which indicates the presence of distribution of relaxation times and hence the relaxation is of non-Debye type and (iii) shifting of peak position towards high frequency region with the rise of temperature which indicates the relaxation process is thermally activated [16]. In the case of localized movement of carriers, the dielectric relaxation is analyzed in the formalisms of electric modulus (M) whereas for the case of long range movement, the resistive and conductive behaviour is often analyzed by electric impedance (Z). Therefore, the combined plot of $M''/M''_{max}$ and $Z''/Z''_{max}$ versus frequency (Fig. 10) can indicate the dominant contributors of dielectric relaxation as the short-range or long-range movement of charge carries. The different peak frequencies for M and Z indicate that the relaxation process is dominated by the short-range movement of charge carriers and deviates from an ideal Debye-type behaviour.

The Fig. 11 shows the $M/M_{max}$ versus $f/f_{max}$ for BNFM-05 ceramics, where $f_{max}$ is the loss peak frequency. For relaxation time to be temperature dependent, all the modulus loss profiles must collapsed into one master curve. As shown in the Fig.11, all peaks related to the oxygen vacancies responses indeed collapse into one master curve at different temperatures from 40 to 200 °C. It clearly suggests that the dynamics of oxygen vacancies occurring at different time scales exhibit the same activation energy. It also indicates that the distribution of its relaxation times doesn't depend on temperature. The frequency region below the maximum value of $M''/M''_{max}$ indicates the range of frequencies for the charge carriers to perform long range hopping from one site to another site. Whereas, frequencies above $M''_{max}$, charge carriers perform short range hopping and these particles are confined to their potential wells. To understand these charge carrier hopping mechanism, ac conductivity has been studied which is discussed in the next section.

Figure 12 shows the frequency dependence of ac conductivity of BNFM-05 ceramics at different temperatures. The response of the material to the applied electric field is described by the ac conductivity. These studies will be useful to investigate the nature of transport mechanism in the present samples. The ac electrical conductivity was calculated by using the eq$^n$ (4).

$$\sigma_{ac} = \varepsilon_0 \varepsilon_r \omega \, tan\delta \qquad \text{------------ (4)}$$

The frequency dependence of ac conductivity obeys Jonscher's power law,

$$\sigma_{ac} = \sigma_0 + A\omega^n = \sigma_0 + \sigma(\omega) \qquad \text{------------ (5)}$$

The parameters '$\sigma_0$' and '$n$' in eq$^n$ (5) are the dc conductivity and frequency exponent respectively. The parameter '$A$' in eq$^n$ (5) is the temperature dependent pre-exponential factor. The exponent n represents the degree of interaction between the mobile ions with the lattice around them [17] whose value also depends on material and temperature [18]. The value of $n = 1$ is meant for Debye behaviour, where the interaction between the neighbouring dipole is negligible. Experimental data points as well as fitted curves to Jonscher's power law [19] have been shown in Fig. 13. The temperature dependence of '$n$' value can suggest the different conduction mechanisms in the materials depending on the various models [20-24]. (1) In the case of quantum mechanical tunneling (QMT) through the barrier separating the two localized sites, the value of '$n$' is independent of temperature (2) '$n$' slightly decrease with increase in temperature in the case of correlated barrier hopping (CBH) (3) '$n$' increases with the increase in temperature in the case of small polaron conduction and (4) '$n$' reaches a minimum followed by a further increase with the increase in the temperature in the case of conduction mechanism of overlapping large polaron tunneling (OLPT). For present ceramics, it was observed that n values increases with increase in temperature. Therefore, the observed electrical transport properties are consistent with the small polaron hopping conduction mechanism and the conduction process is thermally activated.

Figure 14 shows the Arrhenius plots of ac conductivity at different frequencies for BNFM-05. One can distinguish a change in the slope of the electric conductivity plot in vicinity of the magnetic phase transition (Neel temperature, $T_N$) temperature. In region I (below 393 K), conductivity is frequency dependent, but temperature independent. The ac conductivity increases with the increase of frequency due to the increase of hopping of charge carriers. In this region, strong frequency dispersion is observed. Activation energies are found to decrease with the increase in frequency. Low activation energies are observed and attributed to the electronic hopping conduction. In region II (393 K–503 K), conductivity not only depends on temperature, but also on frequency. Contribution of short range of oxygen vacancies to the conductivity is expected in this region. In region III (503 K–573 K), conductivity is frequency independent and temperature dependent. Conductivity is attributed to long range movement of oxygen vacancies or creation of defects. Activation energies increase with the increase in temperature.

To study the relaxation phenomena in the material, the real and imaginary parts of dielectric constant were calculated and plotted against the frequency for all temperatures and are shown in Figs. 15 & 16. A strong dispersion is observed in the real dielectric constant at low frequencies. Low frequency dispersion of BFO is reported to be due to the finite conductivity arising from oxygen vacancies in the BFO ceramics [16]. There are four types of polarization contributions such as (a) electronic, (b) atomic and ionic, (c) dipolar and (d) interfacial polarization. Dipolar polarization contributes in the frequency range of $10^3$–$10^6$ Hz [25]. As the frequency increases its contribution from dipolar polarization decreases because they are unable to follow the electric field in the higher frequency region. Also, the interfacial polarization contributes only in the lower frequency range (~$10^3$ Hz) and hence dielectric constant becomes smaller at higher frequencies because all polarizations do not contribute at high frequencies. The dispersion of dielectric constant with frequency may be attributed to dipole relaxation phenomenon due to Maxwell–Wagner type of interfacial polarization contribution Maxwell–Wagner type interfacial polarization in agreement with Koop's phenomenological theory. The dielectric relaxations give rise to a peak in $\epsilon''$ versus frequency plots at all temperature which shifts to higher frequency with increase in temperature indicating the decrease in dielectric relaxations.

To further confirm the contributions of multiple charge carriers to the dielectric relaxations, the real part of dielectric versus frequency data were fitted to the modified Debye's function that considers the possibility of more than one charge carriers contributing to the dielectric relaxation. [26, 27]. The observed dispersion of the dielectric constant can be modeled using the eq[n] (6).

$$\epsilon' = \epsilon'_\infty + \frac{\epsilon'_o - \epsilon'_\infty}{[1+(\omega\tau)^{2(1-\alpha)}]} \quad \text{-----------(6)}$$

The parameter $\epsilon'$ is the real part of the dielectric constant, $\epsilon'_o$ and $\epsilon'_\infty$ are real part of the dielectric constants at lowest possible frequency (in this case 100 Hz) and higher frequency (at 1 MHz in present case) respectively, $\omega = 2\pi f$ is the angular frequency with f is the applied frequency to the signal of 1 V, τ is the mean relaxation time, and α is the spreading factor of the actual relaxation times about the mean value. Figure 15 depicts the experimentally measured frequency dependence of $\epsilon'$ is analyzed by using eq[n] (7) and the parameters '$\alpha$' and '$\tau$' are enlisted in the Table 3. The experimental data is in good agreement with the calculated data indicating the validity of modified Debye's function with the possibility of more than one ion contributing to the relaxation process.

Figure 17 shows the real part of dielectric ($\epsilon'$) versus temperature for BNFM-05 ceramics for the set of five selected frequencies, viz, 1, 10, 100, 500 and 1000 kHz. The temperature dependence of $\epsilon'$ indicates the two dielectric relaxations in the frequency range of 1–100 kHz. However, only single dielectric relaxation is noted for higher frequencies (500 & 1000 kHz). The dielectric anomaly at the magnetic transition temperature i.e. Neel temperature ($T_N$) has been observed which was predicted on the basis of Landau theory in multiferroic materials on account of magnetoelectric coupling [28]. However, such dielectric anomaly in sintered polycrystalline samples of multiferroic materials at the magnetic transition temperature may not necessarily arise from multiferroicity sample. It may also result from the significant contributions from the interfacial layers such as grain boundaries and grain–electrode interfaces. In order to figure out the cause of the anomalies in dielectric permittivity, equivalent circuit analysis of the impedance spectra has been carried out. The circuit analysis shows the contribution of space charge present at grain boundaries and grain electrode surfaces is negligible at the high frequencies (> 100 KHz). The intrinsic dielectric constant of the grains is equal to the measured value. Therefore, the dielectric anomalies at the high frequencies (> 100 KHz) are intrinsic and could not be due to extrinsic factors such as electrode-sample interface or other interfaces such as grain boundaries. The dielectric anomaly has been observed near $T_N$ which may be attributed to magnetoelectric effect. However, the transition region becomes broad due to oxygen nonstoichiometry and inhomogeneity in the sample due to Bismuth evaporation during the annealing process. The distinct dielectric anomaly for only high (1 MHz) frequency was also observed near ferroelectric transition temperature ($T_c$) because it was broad peak for lower frequencies due to several contributions to polarization as discussed above.

There is shift in the dielectric maxima towards higher temperatures with increase in frequency. This is because as the frequency increases, the charge carriers are not able to align with the fast changing field as a result of which there is decrease in polarization. Consequently, higher energies are required to restore polarization, which requires higher temperature to provide the system with such desired energy. Therefore, higher the frequency, higher is the temperature required.

It is worth mentioning that the dielectric constant increases with the increase in the substitution up to 5 % of co-substitution where the maximum lattice strain exist due to phase competition between both the crystal symmetries such as rhombohedral and orthorhombic phase. With the further increase in the co-substitution it decreases with the decrease in the phase competition due to dominant contribution from the single phase of orthorhombic phase. The final phase as orthorhombic crystal symmetry is centrosymmetric in nature which decreases the dielectric constant. Hence, it has been concluded that the lattice strain drive the phase competition between crystal symmetries and controls the dielectric properties.

4. Conclusions

The equal percentage of both rhombohedral and orthorhombic crystal symmetries is present in the $Bi_{0.95}Nd_{0.05}Fe_{0.95}Mn_{0.05}O_3$ which results in the enhanced strain. It leads to the enhanced dielectric behavior in BNFM-05 sample. The ac conductivity data obeys Jonscher's power law and is consistent with the small polaron hopping conduction mechanism for BNFM-05 sample in the temperature range of 40 to 600 °C. The observed activation energies conclude that electronic hopping, oxygen vacancies movement and creation of lattice defects are the contributors to the ac conductivity. The contribution of multiple charge carriers to the dielectric relaxation have been understood by the modified Debye's function. The dielectric anomaly near $T_N$ may be attributed to the magnetoelectric effect in BNFM-05.

**Acknowledgment**

Mr. Pawan Kumar wishes to acknowledge University Grants Commission, India for providing fellowship during his PhD work.

**Table 1** The goodness of fit ($\chi^2$), crystallite size (D) and lattice parameters (*a*, *b*, *c*) obtained by the Rietveld refinement of XRD patterns of ball milled BFO samples.

| Samples | $\chi^2$ | D | *a = b*, *c* for *R3c* (Å) | *a, b, c* for *Pbnm* (Å) |
|---|---|---|---|---|
| **BNFM-025** | 1.42 | 45.2(1) | 5.5778(6), 13.8317(1) | 5.5327(2), 5.6408(7), 8.0446(9) |
| **BNFM-05** | 1.72 | 28.6(8) | 5.5790(6), 13.8270(1) | 5.5594(6), 5.6328(8), 8.0964(2) |
| **BNFM-075** | 1.25 | 30.0(4) | 5.5743(4), 13.7161(4) | 5.6139(8), 5.6325(4), 8.3984(2) |
| **BNFM-10** | 1.27 | 31.0(7) | 5.5745(2), 13.7725(2) | 5.5550(6), 5.5685(8), 7.9058(6) |

**Table 2** Summary of electrical parameters obtained from fitting of measured data to Jonscher's power law and equivalent circuit model as mentioned in Fig. 7 at various temperatures for BNFM-05 ceramics.

| | Jonscher's power law | | | Grain and Grain boundary parameters | | | | | | |
|---|---|---|---|---|---|---|---|---|---|---|
| Temp (°C) | $\sigma_{dc}$ ($10^{-7}$) | A ($10^{-7}$) | n | $R_s$ (Ω) | $R_g$ (MΩ) | $C_g$ (nF) | $R_{gb}$ (MΩ) | $C_{gb}$ (nF) | CPE ($10^{-10}$) | n |
| 40 | 6.524 | 0.536 | 0.47577 | 58.93 | 20.02 | 0.058 | 2.014 | 0.058 | 7.051 | 0.538 |
| 100 | 300 | 10.371 | 0.53966 | 40.59 | 1.338 | 0.092 | 1.836 | 0.073 | 4.350 | 0.631 |
| 150 | 2000 | 6.518 | 0.56005 | 38.03 | 0.053 | 0.131 | 0.238 | 0.235 | 9.901 | 0.785 |
| 200 | 11400 | 5.829 | 0.61691 | 13.07 | 0.036 | 0.252 | 0.107 | 0.462 | 1.416 | 0.959 |
| 250 | 46900 | 2.161 | 0.72857 | 18.22 | 0.015 | 0.793 | 0.002 | 0.993 | 1.078 | 0.972 |

**Table 3** Spreading factors (α) and relaxation time (τ) determined for BNFM-05 ceramics at five temperatures such as 40, 100, 150, 200 and 250 °C from the analysis of modified Debye's function to the real part of the dielectric constant.

| Temp (°C) | α | τ (μs) |
|---|---|---|
| 40 | 0.520 | 400 |
| 100 | 0.031 | 110 |
| 150 | 0.032 | 20 |
| 200 | 0.279 | 0.48 |
| 250 | 0.631 | 1.013 |

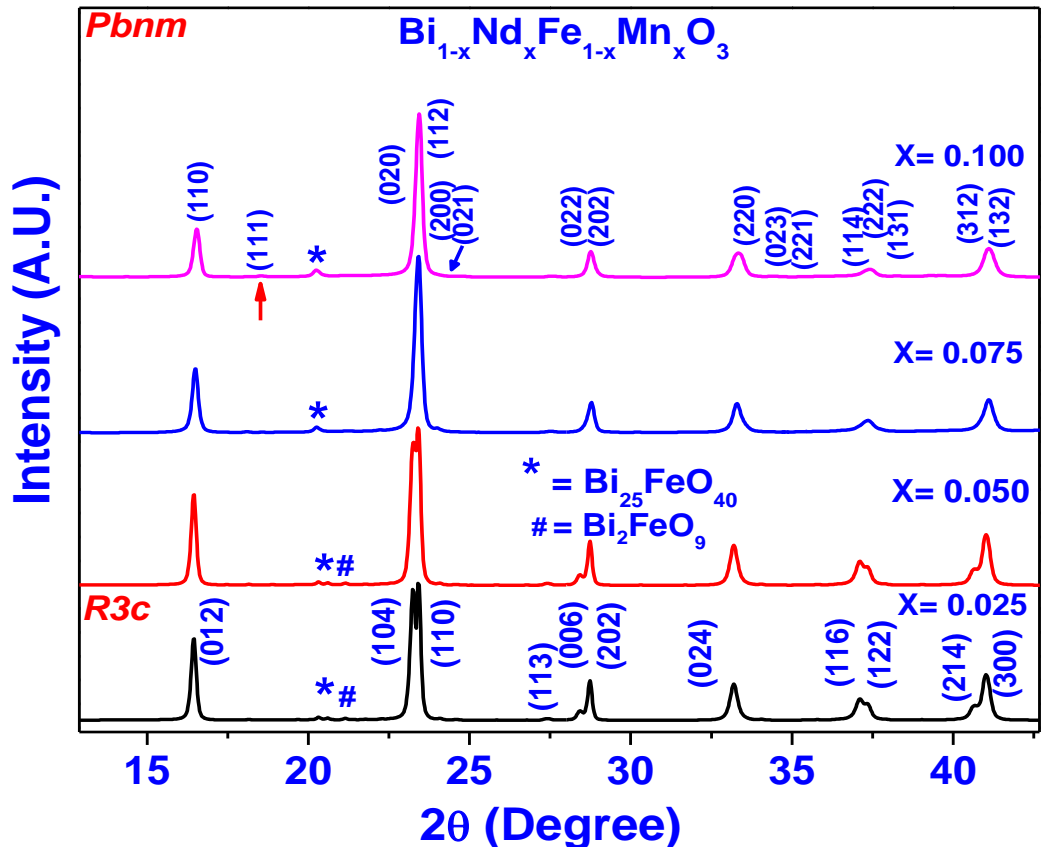

**Fig. 1.** Synchrotron X-Ray diffraction patterns of $Bi_{1-x}Nd_xFe_{1-x}Mn_xO_3$ ceramics for x = 0.025 - 0.100.

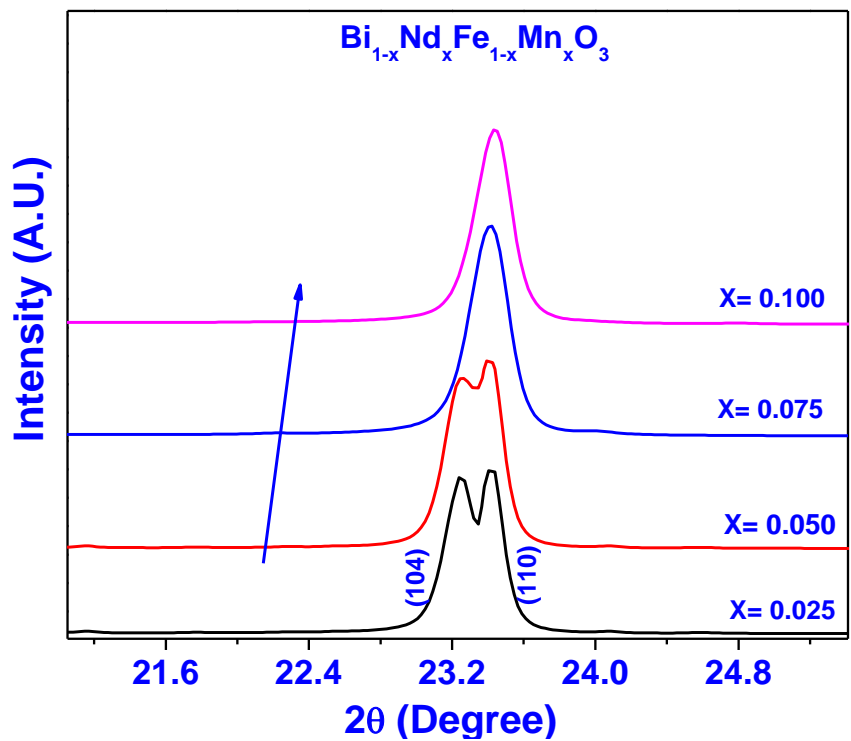

**Fig. 2.** Shift in the Synchrotron X-Ray diffraction patterns of $Bi_{1-x}Nd_xFe_{1-x}Mn_xO_3$ ceramics for x = 0.025 - 0.100.

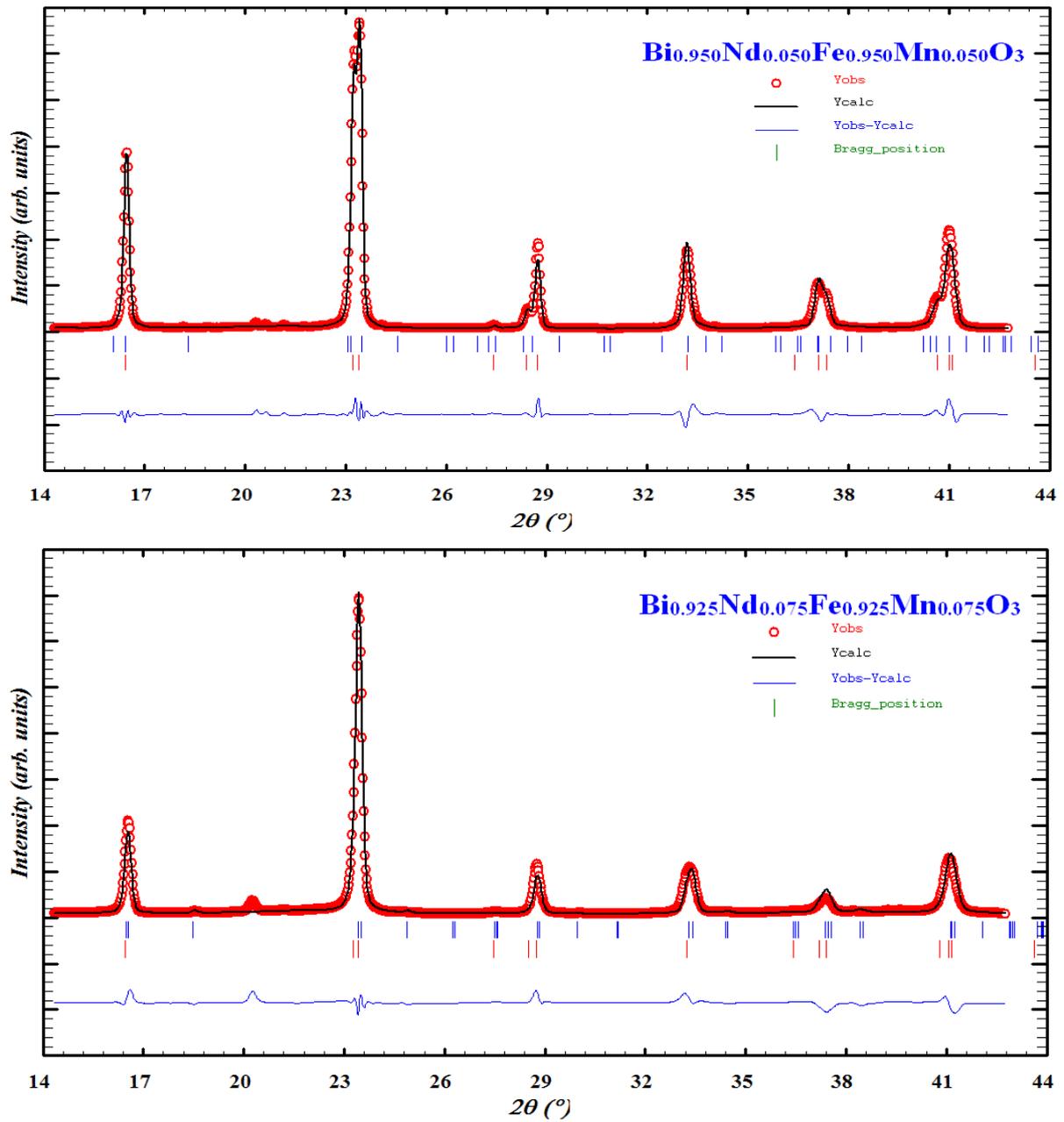

**Fig. 3.** Rietveld refined XRD patterns of $Bi_{1-x}Nd_xFe_{1-x}Mn_xO_3$ samples (x = 0.050 & 0.075). The two rows (1st and 2nd) of Bragg positions are for *Pbnm* and *R3c* space groups respectively.

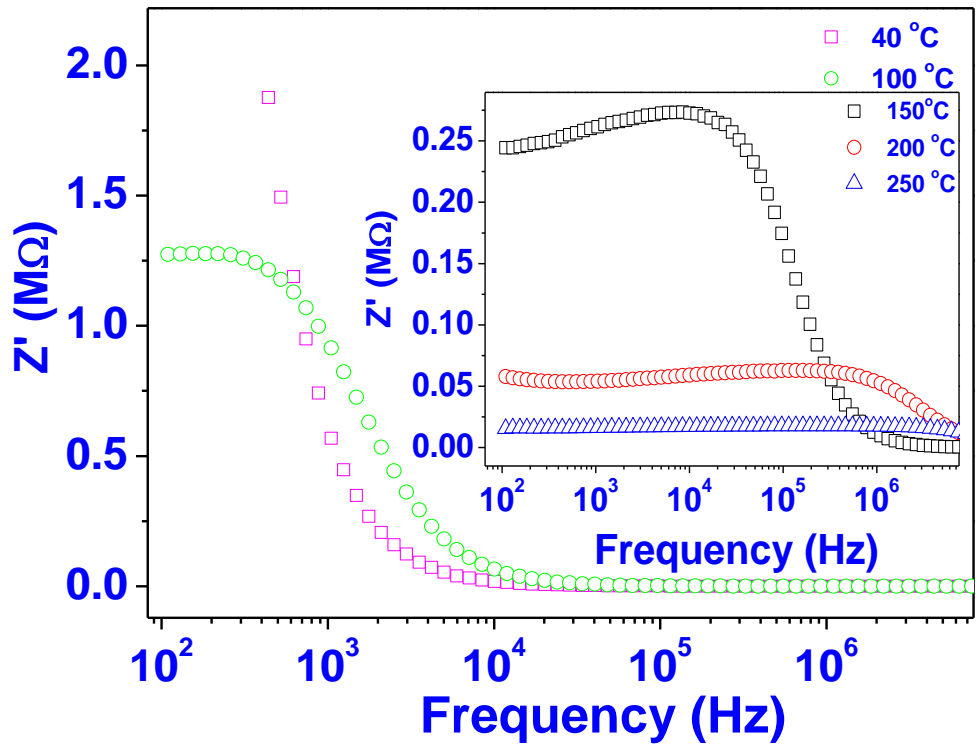

**Fig. 4.** Real part of impedance (Z') versus frequency plot for BNFM-05 ceramics measured at different temperatures.

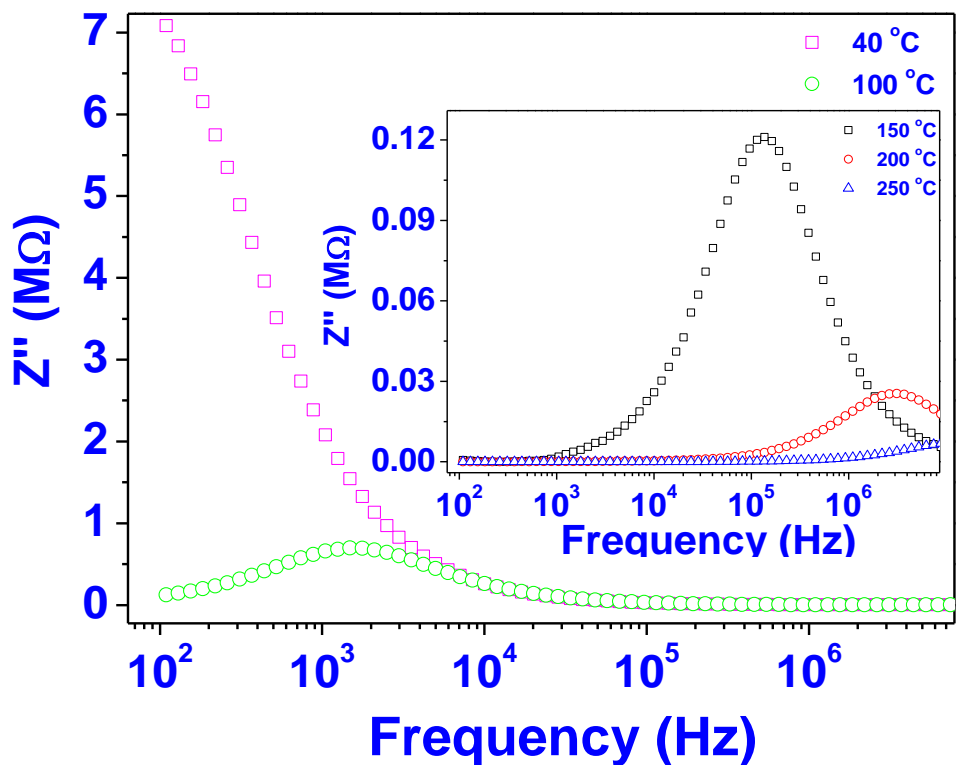

**Fig. 5.** Imaginary part of impedance (Z'') versus frequency plot for BNFM-05 ceramics measured at different temperatures.

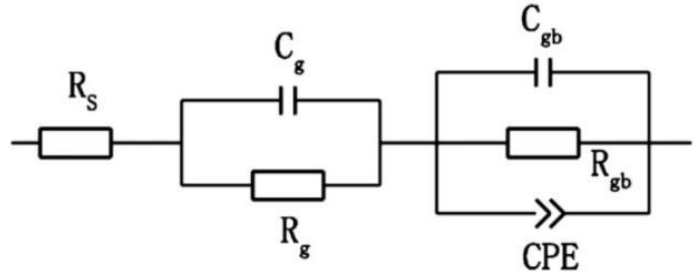

**Fig. 6.** The equivalent circuit model used here in associated with brick layer model.

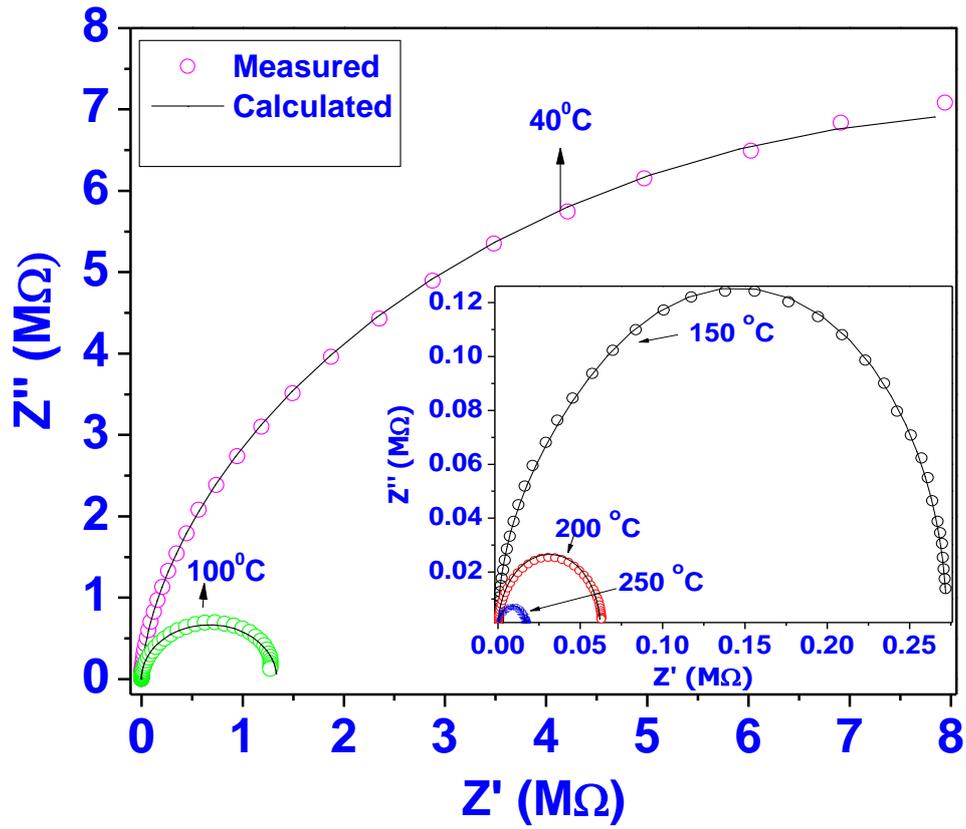

**Fig. 7.** Niquist plot of complex impedance (Z" vs Z') for BNFM-05 ceramics measured at different temperatures.

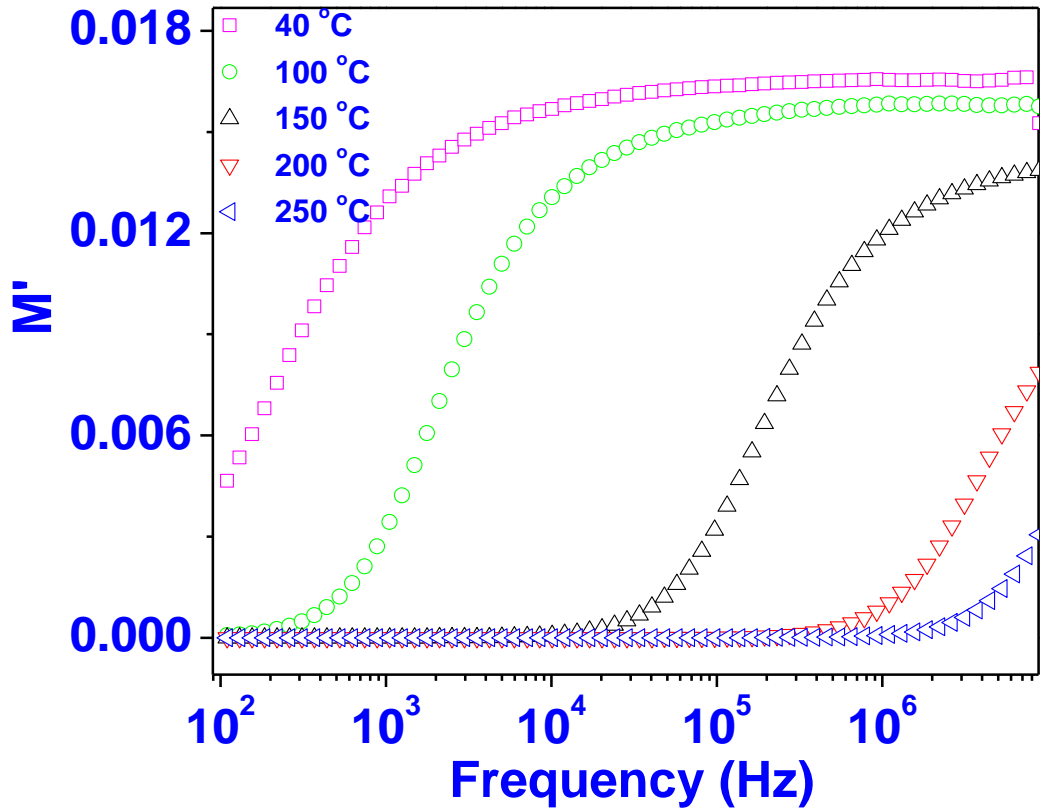

**Fig. 8.** The real part of complex electric modulus (M') versus frequency plot for BNFM-05 ceramics measured at different temperatures.

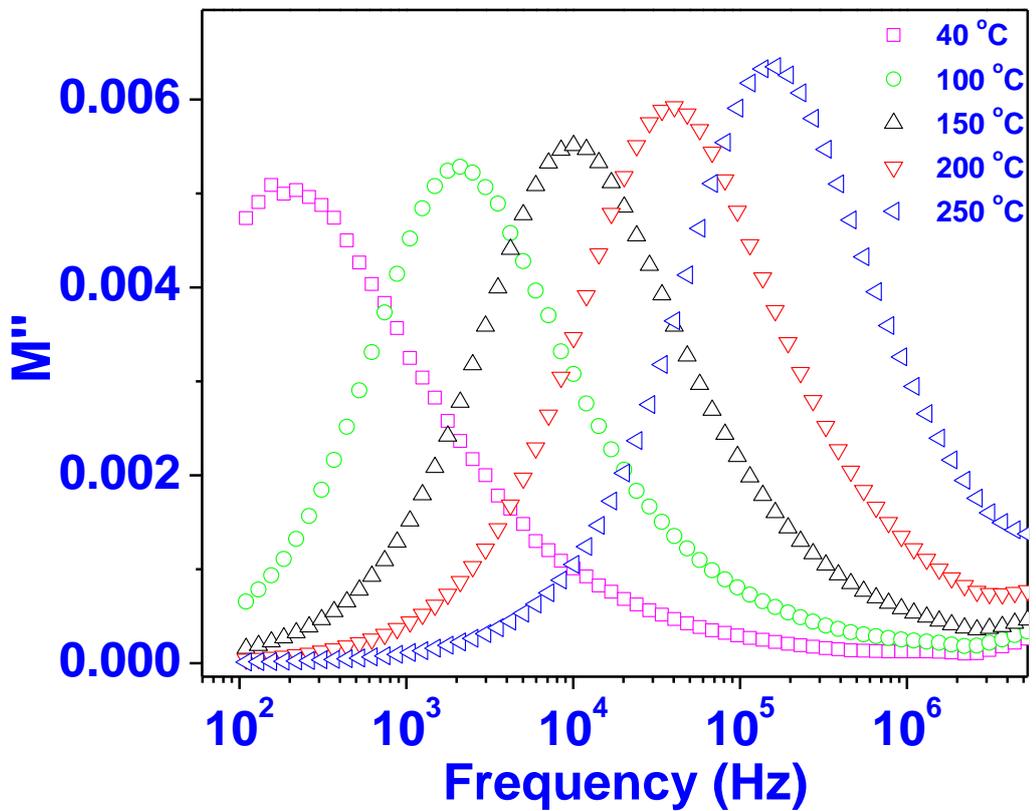

**Fig. 9.** The imaginary part of complex electric modulus (M'') versus frequency plot for BNFM-05 ceramics measured at different temperatures.

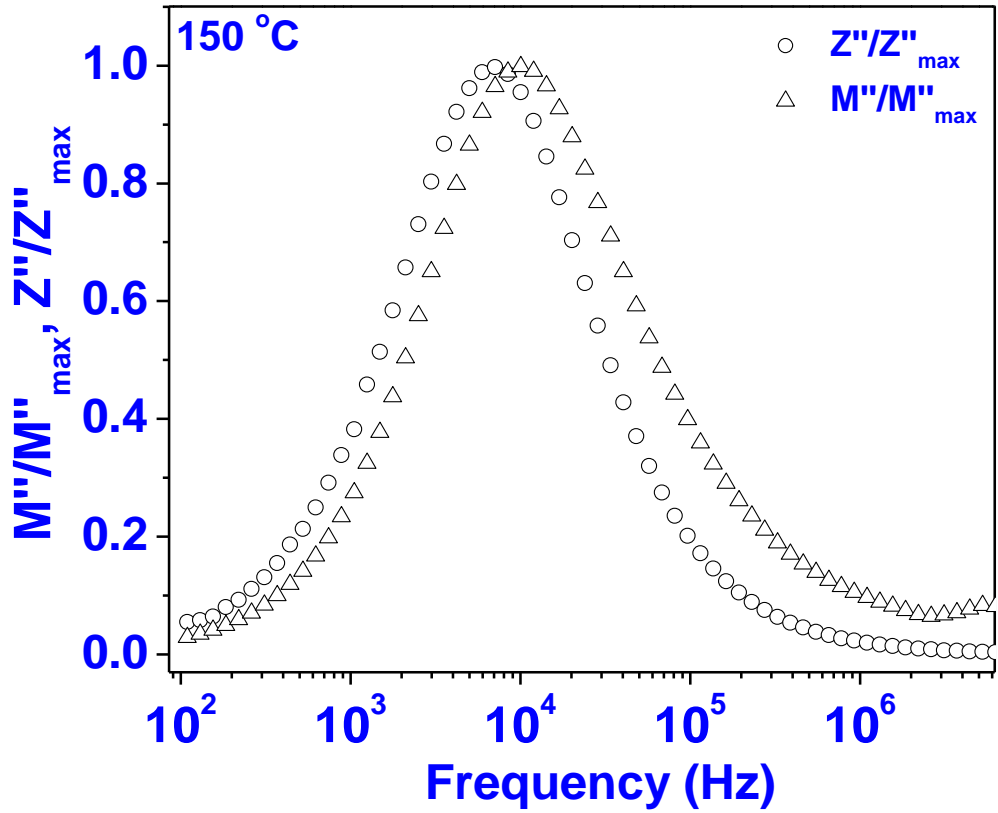

**Fig. 10.** Normalized imaginary part of electric modulus (M''/ M''$_{max}$) and impedance (Z''/ Z''$_{max}$) versus frequency plot for BNFM-05 ceramics.

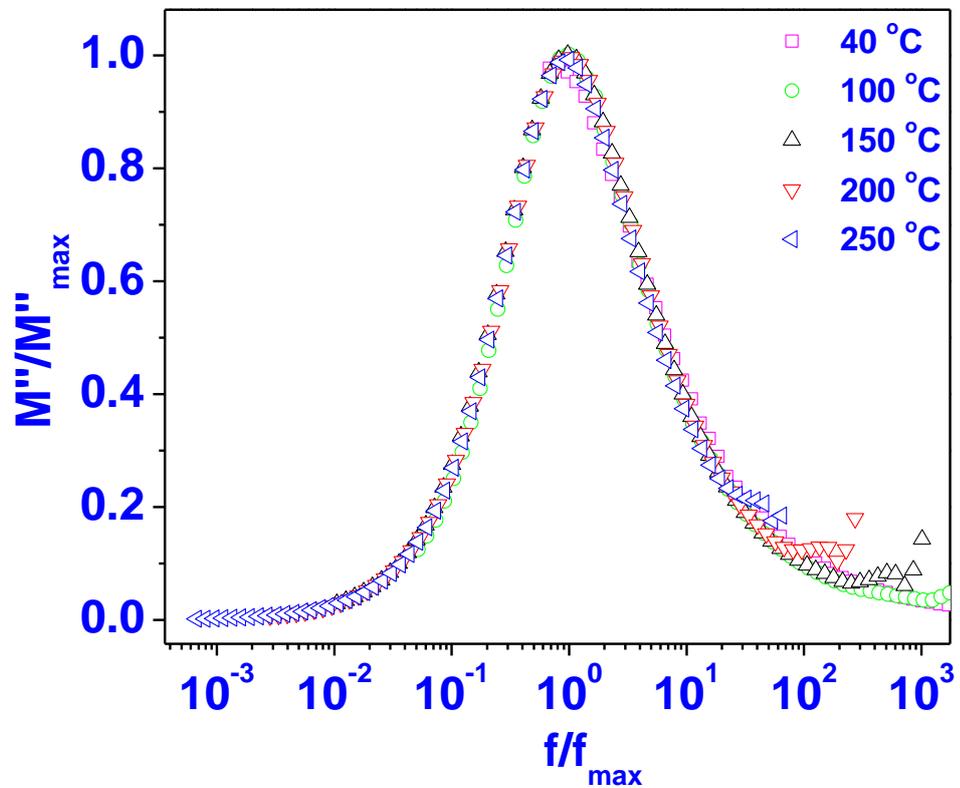

**Fig. 11.** Modulus Master curve (M/M$_{max}$ versus f/f$_{max}$) at various temperatures for BNFM-05 ceramics.

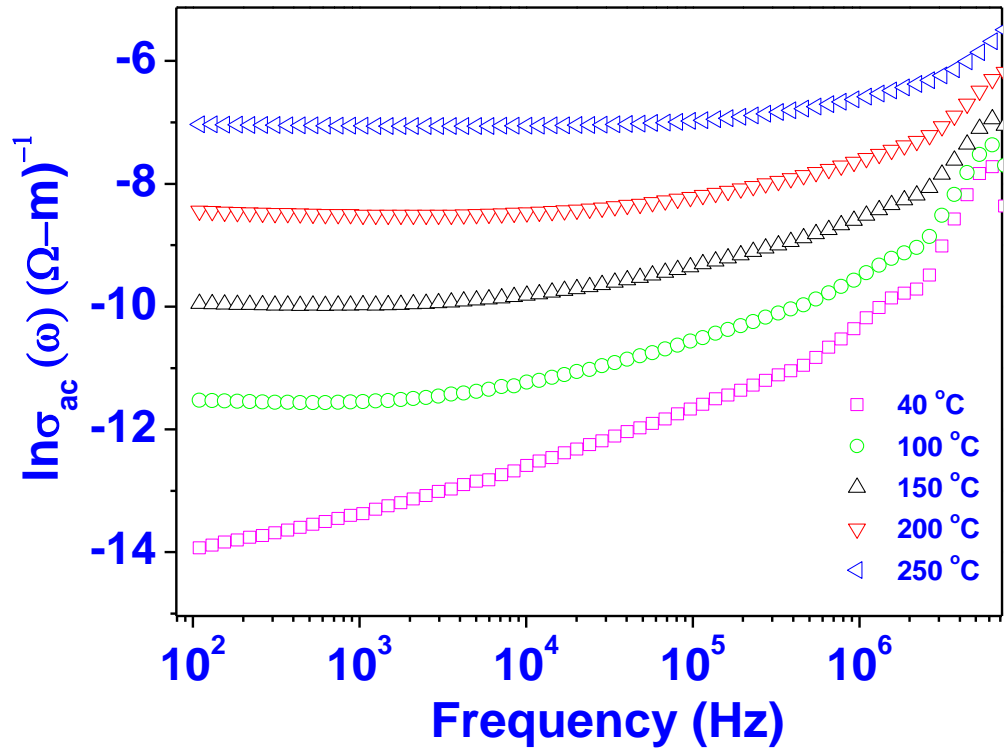

**Fig. 12.** Frequency variation of ac conductivity of BNFM-05 ceramics measured at different temperatures.

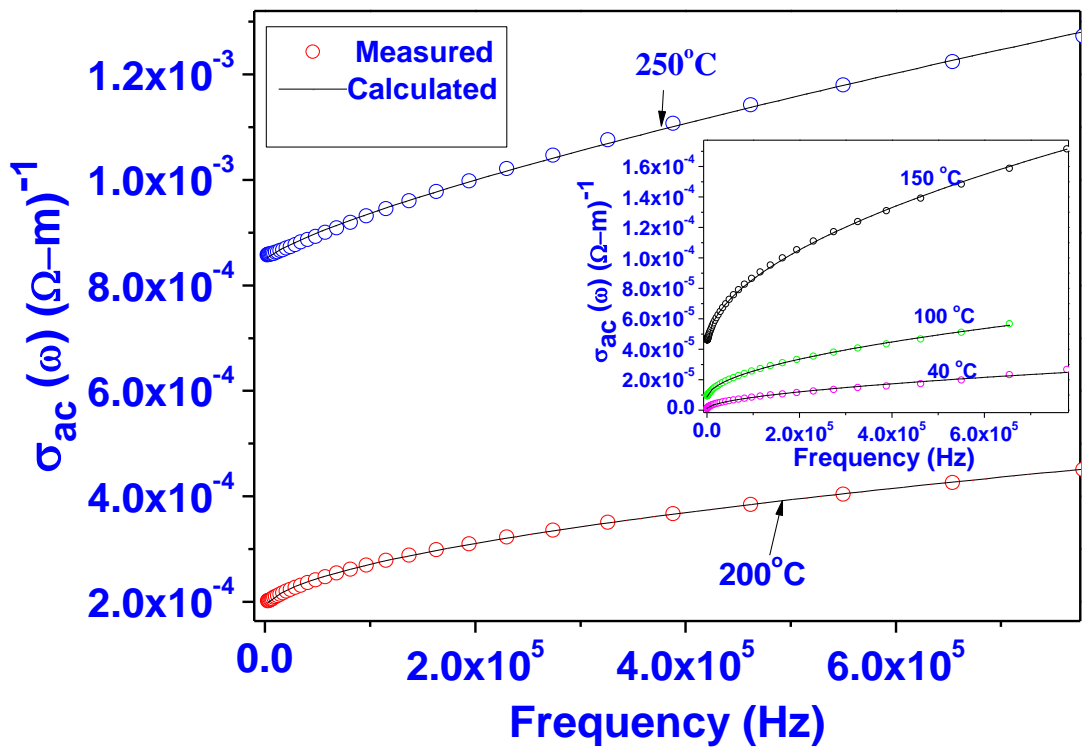

**Fig. 13.** Ac conductivity versus frequency plot fitted with Jonscher's power law for BNFM-05 ceramics measured at different temperatures.

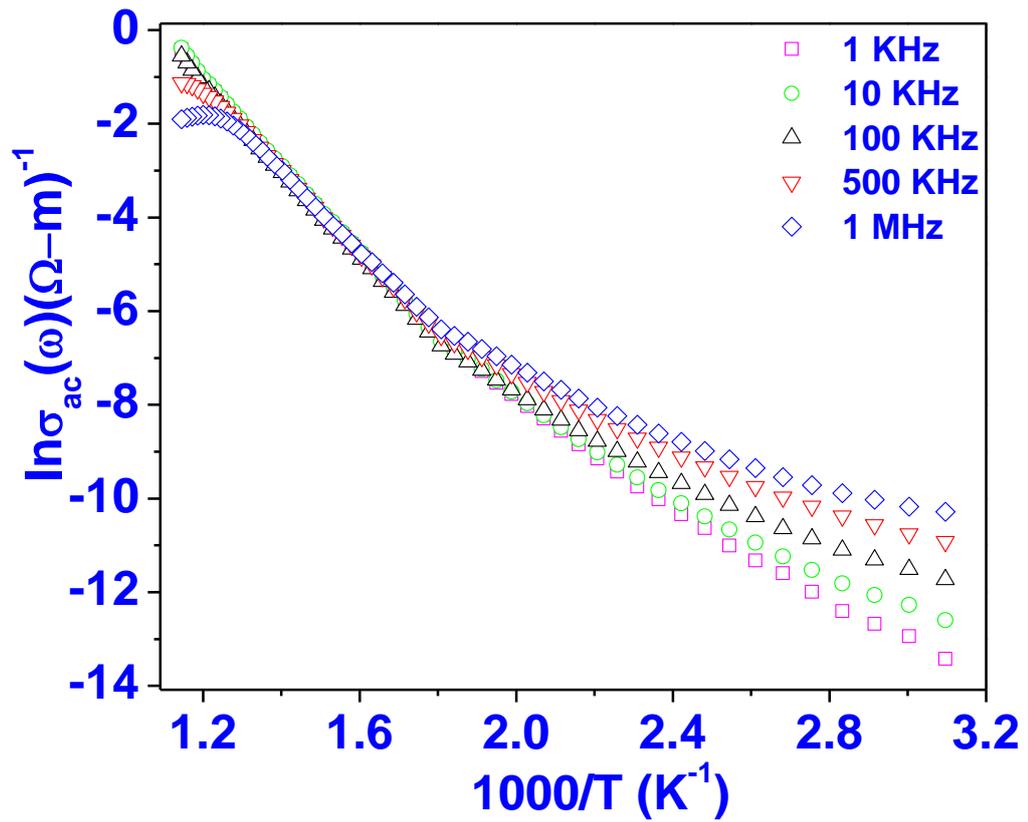

**Fig. 14.** Temperature dependence of ac conductivity for BNFM-05 ceramics measured at different frequencies.

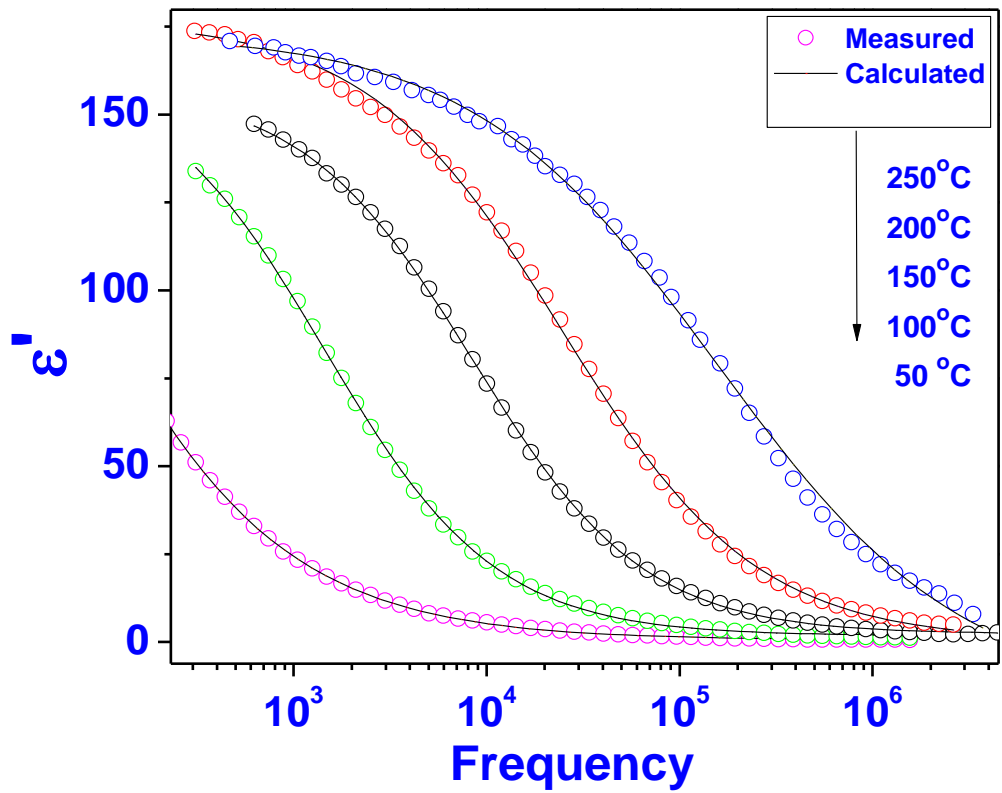

**Fig. 15.** Real part of dielectric constant ($\varepsilon'$) versus frequency plot for BNFM-05 ceramics measured at different frequencies.

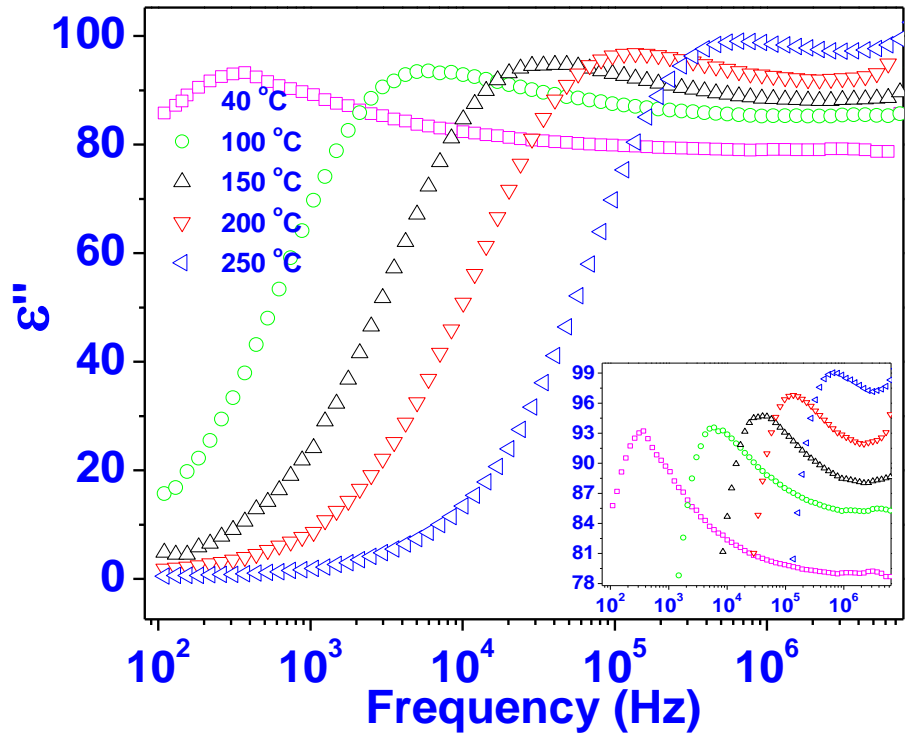

**Fig. 16.** Imaginary part of dielectric constant (**ε''**) versus frequency plot for BNFM-05 ceramics measured at different frequencies. The inset shows the magnified view of the shift in the peak position.

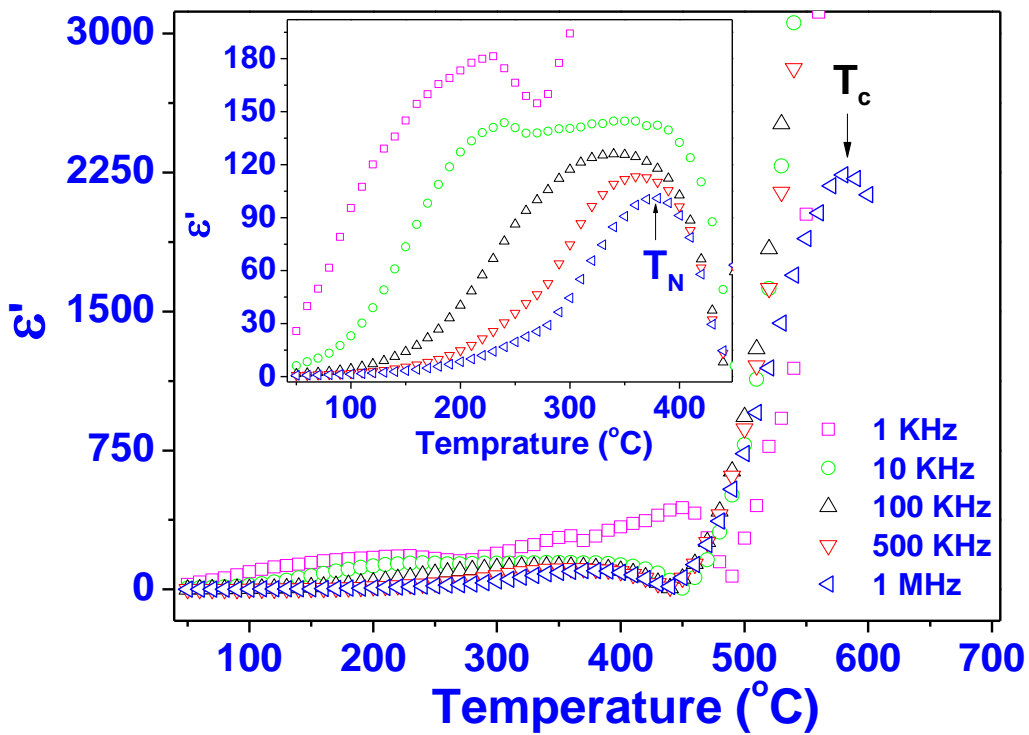

**Fig. 17.** Real part of dielectric constant (**ε'**) versus temperature plot for BNFM-05 ceramics measured at different frequencies. The inset shows the magnified view near magnetic transition temperature (370 °C).